\documentclass[10pt,preprint,letterpaper]{article}\usepackage{opex3}%OE
\usepackage{amsmath,amssymb}
\usepackage{graphicx}
\usepackage{booktabs}

\newcommand\pp{\mathbf{p}}
\newcommand\pwket{|\pp\ \lambda\rangle}
\newcommand\zhat{\mathbf{\hat{z}}}
\newcommand\rhat{\mathbf{\hat{r}}}
\newcommand\xhat{\mathbf{\hat{x}}}
\newcommand\yhat{\mathbf{\hat{y}}}
\newcommand\lhat{\mathbf{\hat{l}}}
\newcommand\z{|p\zhat \ \lambda\rangle}
\newcommand\mz{|-p\zhat \ \lambda\rangle}
\newcommand\pmz{|\pm p\zhat \ \lambda\rangle}
\newcommand\pmzbra{\langle\lambda\ \pm p\zhat|}
\newcommand\zbra{\langle\bar{\lambda}\ p\zhat|}
\newcommand\mzbra{\langle\bar{\lambda}\ -p\zhat|}
\begin{document}
\title{Forward and backward helicity scattering coefficients for systems with discrete rotational symmetry}
\author{Ivan Fernandez-Corbaton$^*$}
\address{ARC Center for Engineered Quantum Systems, Macquarie University, North Ryde, New South Wales 2109, Australia}%OE
%\affiliation{ARC Center for Engineered Quantum Systems, Macquarie University, North Ryde, New South Wales 2109, Australia}%RT
\email{$^*$ivan.fernandez-corbaton@mq.edu.au}
\begin{abstract}
	The forward and backward scattering off linear systems with discrete rotational symmetries $R_z(2\pi/n)$ with $n\ge 3$ are shown to be restricted by symmetry reasons. Along the symmetry axis, forward scattering can only be helicity preserving and backward scattering can only be helicity flipping. These restrictions do not exist for $n<3$. If, in addition to the $n\ge 3$ discrete rotational symmetry, the system has duality symmetry (obeys the helicity conservation law), it will exhibit zero backscattering. The results pinpoint the underlying symmetry reasons for some notable scattering properties of $R_z(2\pi/4)$ symmetric systems that have been reported in the metamaterials and radar literature. Applications to planar metamaterials and solar cells are briefly discussed.
\end{abstract}
%\maketitle%RT

%\bibliographystyle{osajnl}%OE
%\bibliography{ivanfcrefs}%OE

%OE

Understanding the interaction of electromagnetic radiation with matter is crucial for our technological development. So far, such understanding has allowed applications like radiotherapy, GPS, the high speed internet and solar cells. For the future, we envision millimeter sized labs, nanorobots controlled with light and devices for the exquisite manipulation of electromagnetic fields. This wide field of applications benefits from improvements of and additions to the set of tools that we use to study light-matter interactions.

One of the most powerful frameworks at our disposal is that of symmetries and conservation laws. Its application has produced many advances in physics, chemistry and other disciplines. When applied to light matter interactions it allows to understand, for example, the selection rules for atomic and molecular excitation, the Zeeman and Stark effects, or the scattering off a spherical target.

In this article, I will use symmetries and conservation laws to study the electromagnetic forward and backward scattering properties of linear systems with discrete rotational symmetries $R_z(2\pi/n)$ for $n=1,2,3,\ldots$ . I will show that the scattering coefficients are restricted for systems with symmetries of degree $n\ge3$: Along the axis of symmetry, forward scattering can only be helicity preserving while backward scattering can only be helicity flipping. These restrictions do not exist for systems with symmetries of degree $n=2$ or the trivial $n=1$. These results depend only on the discrete rotational symmetry properties of the scatterer. I will also show that, if in addition to the discrete rotational symmetry of degree $n\ge3$, the system has electromagnetic duality symmetry, it will exhibit zero backscattering. These results pinpoint the symmetry reasons for some notable effects in the scattering off arrays \cite{Decker2010}, isolated scatterers \cite{Kaschke2012}, and in the context of targets invisible to radar \cite{Lindell2009,Karilainen2012}. In all these cases, the systems have discrete $2\pi/4$ rotational symmetry. The analysis contained in this article shows that all these effects must also exist in systems with $2\pi/n$ discrete rotational symmetries as long as $n\ge3$.

I will begin with a brief introduction to the electromagnetic duality symmetry and its associated helicity conservation law, which can be used for the study of electromagnetic scattering by material systems. I will then concentrate in the special case relevant for this article: The incident field is a single plane wave and only the forward and backward scattering directions are of interest. I will then derive the aforementioned results and link them to the polarization transformation and zero backscattering properties of some systems reported in the metamaterials and radar literature. Finally, I will comment on some possible applications of the results.

\begin{figure}[ht]
\begin{center}
	\includegraphics[width=0.75\linewidth]{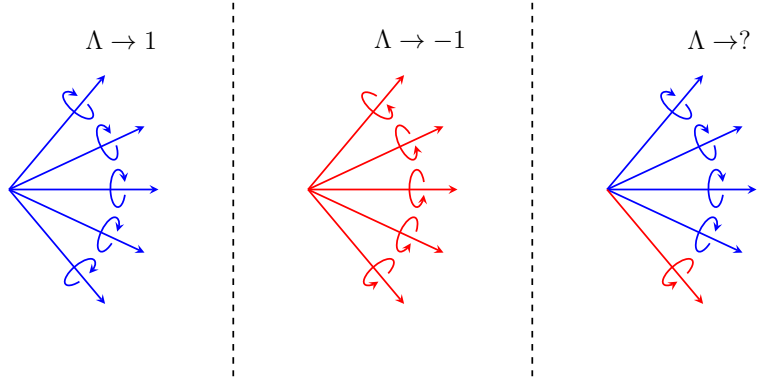}
\end{center}
\caption{A field composed by the superposition of five plane waves has a well defined helicity equal to one if all the plane waves have left handed polarization (left panel), equal to minus one if they all have right handed polarization (central panel) and does not have a well defined helicity, i.e. it is not an eigenstate of the helicity operator, if all the plane waves do not have the same handedness (right panel).}
\label{fig:helicity}
\end{figure}

Electromagnetic duality is a transformation that mixes electric and magnetic fields by means of a real angle $\theta$. In units of $\epsilon_0=\mu_0=1$, its expression is \cite[chap. 6.11]{Jackson1998}: 
\begin{equation}
\label{eq:gendual}
\begin{split}
\mathbf{E}&\rightarrow \mathbf{E}_\theta=\mathbf{E}\cos\theta  - \mathbf{H}\sin\theta , \\
\mathbf{H}&\rightarrow \mathbf{H}_\theta=\mathbf{E}\sin\theta +  \mathbf{H}\cos\theta .
\end{split}
\end{equation}
In vacuum, Eq. (\ref{eq:gendual}) is a symmetry of Maxwell's equations: If the electromagnetic field  $(\mathbf{E},\mathbf{H})$ is a solution of the free space Maxwell's equations, then the field $(\mathbf{E}_\theta,\mathbf{H}_\theta)$ is also a solution for any value of $\theta$. In the 1960's, Calkin \cite{Calkin1965} and Zwanziger \cite{Zwanziger1968} showed that helicity was the conserved quantity related to such symmetry. In other words, helicity and duality have the same relationship as linear momentum and translations or angular momentum and rotations. The momentum $\mathbf{P}$ and angular momentum $\mathbf{J}$ operators generate translations and rotations. The helicity operator $\Lambda=\mathbf{J}\cdot\mathbf{P}/|\mathbf{P}|$ generates the duality transformation. For the transverse electromagnetic field $\Lambda$ has two eigenvalues $\pm1$, and its eigenvectors can be taken to be the Riemann-Silberstein combinations $\mathbf{E}\pm i\mathbf{H}$ \cite{Birula1994,Birula1996,FerCor2013}. It is possible to intuitively understand the meaning of helicity when considering the momentum space decomposition of a general field, that is, as a superposition of plane waves. In this representation, helicity is related to the handedness of the polarization of each and every plane wave. The helicity of a field is well defined, i.e. the field is an eigenstate of $\Lambda$, only when all the plane waves have the same handedness with respect to their momentum vector, including both propagating and evanescent components. Figure \ref{fig:helicity}(a) illustrates this.

\begin{figure}[h]
	\begin{center}
\includegraphics[width=0.75\linewidth]{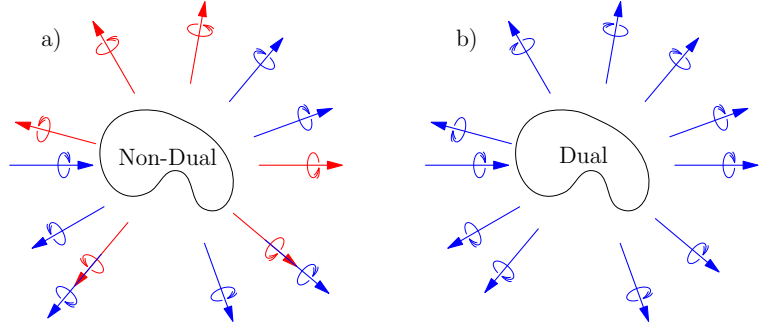}
\caption{(a) The helicity of an electromagnetic field is not preserved after interaction with a non-dual symmetric object. An incoming field with well defined helicity, in this case a single plane wave of definite polarization handedness (blue), produces a scattered field that contains components of the opposite helicity (red). The helicity of the scattered field in panel (a) is not well defined because it contains plane waves of different helicities. (b) Helicity preservation after interaction with a dual symmetric object. The helicity of the scattered field is well defined an equal to the helicity of the input field.} 
\label{fig:dnd}
\end{center}
\end{figure}

In general, the presence of charges breaks the symmetry of the equations: A solution $(\mathbf{E},\mathbf{H})$ does not result in a new solution when transformed as in Eq. (\ref{eq:gendual}). As a consequence, the interaction with matter typically produces components of changed helicity (see Fig. \ref{fig:dnd}). Nevertheless, the restoration of the symmetry is possible in both the macroscopic equations and the dipolar scattering approximations when precise conditions are met \cite{Lindell2009,Karilainen2012,Zambrana2013b,FerCor2012p,FerCor2013}. Recent work shows that the relationship between helicity and duality can be used as a theoretical and experimental tool for the study of light matter interactions \cite{FerCor2012b,FerCor2012p,Zambrana2013,FerCor2012c,Zambrana2013b,FerCor2013}. 

Figure \ref{fig:dnd} depicts the general situation: A non dual symmetric system will not preserve the helicity of the fields that interact with it. In this article, I will deal with a particular case.

Suppose now that we do not care about the whole scattered field and that we are only interested in the forward and backward scattered components resulting from the interaction of a single incoming plane wave with a material system. Obviously, if the system is dual, helicity will be preserved in this reduced set of modes. On the other hand we cannot make a general statement about whether a non-dual system will or will not preserve helicity in the directions that we care about. The only thing that we can say for a general non-dual system is that there exist at least a pair of input and scattered directions for which helicity is not preserved. I will now show that, independently of their duality properties, we can make definite statements about helicity preservation/change for the forward and backward scattering directions of systems with discrete rotational symmetries. This is possible because in forward and backward scattering, contrary to the more general case shown in Fig. \ref{fig:dnd}, the momenta of the input and scattered plane waves are parallel or anti-parallel to a unique axis (which I will take to be the $z$ axis). In this case, helicity and angular momentum along such axis ($J_z$) are related and this relationship connects helicity with the rotational properties of the system. In general, though, helicity and angular momentum are two decoupled properties of the field \cite{FerCor2012p,FerCor2012b}. This decoupling can be appreciated by the impossibility of associating a definite angular momentum to the eigenstates of helicity depicted in the left and central panels of Fig. \ref{fig:helicity}. It is also reflected in the fact that beams of well defined $J_z$, like Bessel beams or multipoles can be obtained as arbitrary linear combinations of two modes of well defined helicity \cite[app. B]{FerCor2012b}.

Two notes about notation and assumptions before getting into it.

In the derivations, I will denote a plane wave of defined momentum $\pp$ and helicity $\lambda$ by $\pwket$. Since we only care about forward and backward scattering, I will only need plane waves with momentum parallel or anti parallel to the positive $z$ axis $\pmz$, where $p=|\pp|$. Table \ref{tab:helreal} contains the explicit expressions of $\pmz$ in the real representation of monochromatic fields. Note how plane waves of opposite helicity and momentum have the same real space polarization vector. That helicity and real space polarization are two different concepts is very evident in multipoles or Bessel beams of well defined helicity \cite[sec. VI-A]{FerCor2012b}, which have complicated real space polarization maps.

\begin{table}[h]
\caption{\label{tab:helreal} Expressions for monochromatic plane waves of well defined helicity ($\lambda=\pm 1$) in the real space representation of electromagnetic fields. The momentum of the plane waves is either parallel or anti-parallel to the $z$ axis $\pm p\zhat$. The real space polarization vectors in the expressions are $\lhat=(\xhat+i\yhat)/\sqrt{2}$ and $\rhat=(\xhat-i\yhat)/\sqrt{2}$. To have the same handedness with respect to its momentum vector, the real space polarization vector must change when the momentum changes sign.}
\begin{center}\begin{tabular}{rcc} \toprule
	&	$\lambda=1$ & $\lambda=-1$ \\ \midrule
$p\zhat\textrm{\hspace{0.28cm}}$&$-\lhat\exp(i(pz-\omega t))$ & $\rhat\exp(i(pz-\omega t))$\\
$-p\zhat\textrm{\hspace{0.3cm}}$   &$\rhat\exp(i(-pz-\omega t))$ & $-\lhat\exp(i(-pz-\omega t))$\\
	\bottomrule
\end{tabular}\end{center}
\end{table}

The helicity dependent forward ($\tau_f^{\lambda\bar{\lambda}}$) and backward ($\tau_b^{\lambda\bar{\lambda}}$) scattering coefficients are the following matrix elements of the scattering operator $S$ which models the interaction of the incoming field with the material system:
\begin{equation}
	\tau_f^{\lambda\bar{\lambda}}=\zbra S \z,\ \tau_b^{\lambda\bar{\lambda}}=\mzbra S \z.
\end{equation}
$\lambda(\bar{\lambda})$ is the helicity of the input(scattered) plane waves. I will assume that $S$ is linear and preserves the frequency of the fields ($\omega=|\pp|=p$ in units of $c=1$). 

I now establish the relationship between helicity and angular momentum for $\pmz$ states and obtain their transformation properties under rotations along the $z$ axis. The definition of $\pwket$ as helicity eigenstates means that  
\begin{equation}
	\label{eq:lambda}
	\Lambda\pmz=\lambda\pmz.
\end{equation}
Let me now expand the helicity operator in Eq. (\ref{eq:lambda}):
\begin{equation}
	\label{eq:jz}
	\Lambda\pmz=\frac{\mathbf{J}\cdot\mathbf{P}}{|\mathbf{P}|}\pmz=\frac{\sum_{i=1}^3 J_iP_i}{\sqrt{\sum_{i=1}^3 {P_i}^2}}\pmz=\pm J_z\pmz,\\
\end{equation}
where the last equality follows from applying the $P_i$ operators to the plane waves with momentum $\pp=[0,0,|\pp|=p]$. From Eqs. (\ref{eq:lambda}) and (\ref{eq:jz}), it follows that the plane wave helicity eigenstates $\pmz$ are also angular momentum ($J_z$) eigenstates: 
\begin{equation}
J_z\pmz=\pm\lambda\pmz.
\end{equation}

In the case of $\z$, the eigenvalues of $\Lambda$ and $J_z$ coincide, while for $\mz$, they have opposite sign. Accordingly, the properties of $\pmz$ under rotations along the $z$ axis are:
\begin{equation}
	\label{eq:ralpha}
	R_z(\alpha)\pmz=\exp(\mp i\alpha\lambda)\pmz.\\
\end{equation}
I will also need their hermitian conjugate version:
\begin{equation}
	\label{eq:ralphaherm}
	\begin{split}
		&\pmzbra R_z(\alpha)^{-1}=\left(R_z(\alpha)^{-\dagger}\pmz\right)^{\dagger}\\
	&=\left(R_z(\alpha)\pmz\right)^{\dagger}=\exp(\pm i\alpha\lambda)\pmzbra,
	\end{split}
\end{equation}
where I have used that rotations are unitary transformations $R^{-1}(\alpha)=R^\dagger(\alpha)\implies R^{-\dagger}(\alpha)=R(\alpha)$. The relationship between helicity and angular momentum for single plane waves can be checked by applying a rotation to the expressions in Table \ref{tab:helreal}. 

\begin{figure}[h]
\begin{center}
	\includegraphics[width=0.8\linewidth]{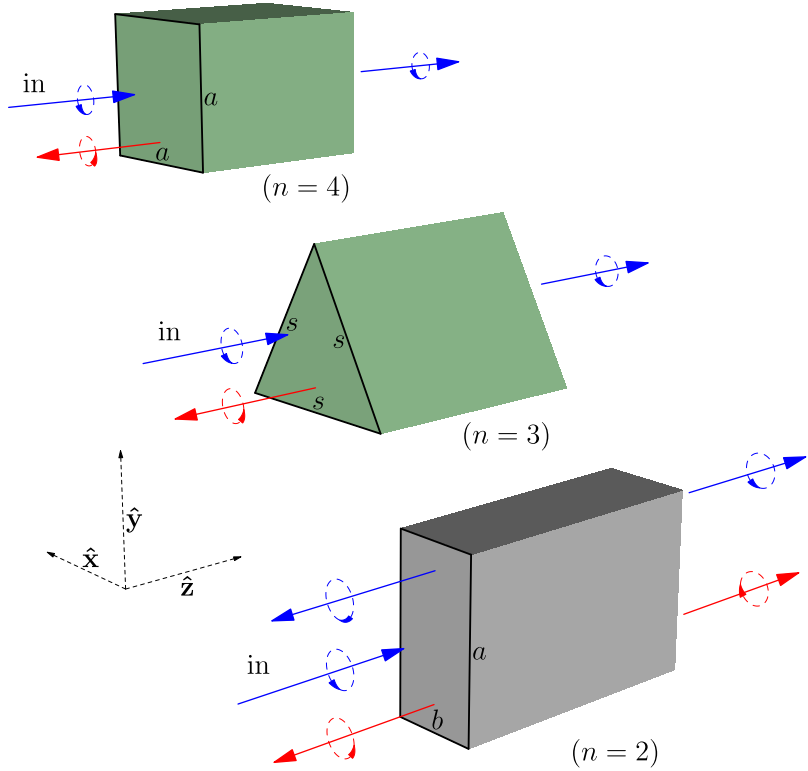}
\end{center}
\caption{Forward and backward scattering produced by input plane waves of well defined helicity impinging on structures with discrete rotational symmetries $R_z(2\pi/n)$: Gray rectangular prism ($n=2$), green equilateral triangular prism ($n=3$) and green square prism ($n=4$). Plane waves of positive helicity (left handed polarization) are blue. Plane waves of negative helicity (right handed polarization) are red. The input plane waves, labeled as ``in'', have positive helicity and momentum aligned with the positive $z$ axis. The text shows that for $n\ge 3$, the forward scattering can only contain components with the same helicity as the input and the backward scattering can only contain components with helicity opposite to the input one. In the figure, this is reflected in the restricted helicity components drawn in forward and backward scattering for the square ($n=4$) and triangular ($n=3$) prisms. No such restrictions apply to the rectangular prism ($n=2$): Any helicity is allowed both in forward and in backward scattering. If, besides the discrete rotational symmetry, the scatterers had also electromagnetic duality symmetry, all the red right handed plane waves will disappear from the picture because duality enforces helicity preservation in all scattering directions. Therefore, dual objects with discrete rotational symmetries with $n\ge 3$ will exhibit zero backscattering.\label{fig:nyaca}} 
\end{figure}

I will now assume that the scattering system has a discrete rotational symmetry  $2\pi/n$ along the $z$ axis. For example, for $n=2,3,4$, this corresponds to the symmetry of a prism whose base is a rectangle, a triangle with equal sides, or a square, respectively (see the objects in  Fig. \ref{fig:nyaca}). What this symmetry means for the scattering operator is that $S$ is invariant under a transformation by $R_z(2\pi/n)$:
\begin{equation}
\label{eq:rsr}
	R_z^{-1}(2\pi/n)SR_z(2\pi/n)=S.
\end{equation}
All is now ready. Let me first study the forward scattering coefficient $\tau_f^{\lambda\bar{\lambda}}$. I will write the reference to the formulas that are needed for each step on top of the corresponding equality:
\begin{equation}
\label{eq:f}
	\begin{split}
		\tau_f^{\lambda\bar{\lambda}}&=\zbra S \z\stackrel{\ref{eq:rsr}}{=}\zbra R_z^{-1}(2\pi/n)SR_z(2\pi/n) \z\\
		&\stackrel{\ref{eq:ralpha},\ref{eq:ralphaherm}}{=}\exp(-i(\lambda-\bar{\lambda})\frac{2\pi}{n})\zbra S \z\\
		&=\exp(-i(\lambda-\bar{\lambda})\frac{2\pi}{n})\tau_f^{\lambda\bar{\lambda}}.
	\end{split}
\end{equation}
For helicity preserving scattering $\lambda=\bar{\lambda}$, Eq. (\ref{eq:f}) results in the trivial $\tau_f^{\lambda\bar{\lambda}}=\tau_f^{\lambda\bar{\lambda}}$, which contains no information. But, for helicity changing scattering $\lambda=-\bar{\lambda}$, there are only two ways to meet $\tau_f^{\lambda\bar{\lambda}}=\tau_f^{\lambda\bar{\lambda}}\exp(\pm i 4\pi/n)$. One is that $\tau_f^{\lambda\bar{\lambda}}=0$, and the other is that there exist an integer $k$ such that:
\begin{equation}
\label{eq:ff}
	\frac{4\pi}{n}=2\pi k \implies \frac{2}{n}=k,
\end{equation}
This second way is only possible if $n=1$ or $n=2$. It can not happen for $n\ge 3$, which then forces $\tau_f^{\lambda\bar{\lambda}}=0$. This means that there is no component of changed helicity in the forward scattering direction of a system with a discrete rotational symmetry $R_z(2\pi/n)$ with $n\ge 3$. The forward scattering direction can only contain the preserved helicity component. Note that the derivation does not involve the duality properties of the scatterer, i.e. its general helicity preservation properties.

It is now the turn of the backward scattering coefficient. 
\begin{equation}
\begin{split}
	\label{eq:back}
	\tau_b^{\lambda\bar{\lambda}}&=\mzbra S \z\\
								 &\stackrel{\ref{eq:rsr}}{=}\mzbra R_z^{-1}(2\pi/n)SR_z(2\pi/n) \z\\
			   &\stackrel{\ref{eq:ralpha},\ref{eq:ralphaherm}}{=}\exp(-i(\lambda+\bar{\lambda})\frac{2\pi}{n})\mzbra S \z\\
			   &=\exp(-i(\lambda+\bar{\lambda})\frac{2\pi}{n})\tau_b^{\lambda\bar{\lambda}}.
	\end{split}
\end{equation}
The situation is now reversed with respect to the helicity. For helicity flipping backscattering $\lambda=-\bar{\lambda}$, the result is trivial $\tau_b^{\lambda\bar{\lambda}}=\tau_b^{\lambda\bar{\lambda}}$. For helicity preserving backscattering $\lambda=\bar{\lambda}$, you can go through the same arguments as before to conclude that, there is no preserved helicity component in the backward scattering direction of a system with a discrete rotational symmetry $R_z(2\pi/n)$ with $n\ge3$. The backward scattering direction can only contain the changed helicity component.

In summary, for $n=1,2$, the two helicities are possible in both forward and backward scattering. For $n\ge3$, forward scattering can only be helicity preserving and backward scattering can only be helicity flipping. Figure \ref{fig:nyaca} illustrates these results. This analysis is only valid for the forward and backward scattering directions. 

Two very clear examples for $n=4$ can be found in \cite{Decker2010} and \cite{Kaschke2012}. In \cite{Decker2010} the authors design an array of split ring resonators which has $R_z(2\pi/4)$ symmetry. Their forward scattering results show very small ($<10^{-5}$) conversion ratios between the two circular polarizations. 
In \cite{Kaschke2012}, the authors study the forward and backward scattering properties of their design, which consist of four gold helices set in a square, specifically arranged to have $R_z(2\pi/4)$ symmetry. This structure is compared with a single helix, which lacks rotational symmetries (except the trivial $n=1$). The authors analyze several cases where the helices have different number of turns and use both a lossless and a lossy model for the response of the gold. Their results show zero circular polarization conversion ratio in forward scattering for the $n=4$ structure, regardless of the number of helix turns and the loss model. The single helix shows non-zero conversion in both lossless and lossy cases.

These results from the two references are consistent with the restrictions obtained in Eqs. (\ref{eq:f}) and (\ref{eq:ff}): Helicity cannot change in forward scattering when $n\ge 3$. Helicity preservation in forward scattering translates in preservation of the real space circular polarizations ($\rhat,\lhat$) (see Table \ref{tab:helreal}).

The backward scattering analysis in \cite{Kaschke2012} shows preservation of the real space circular polarizations ($\rhat,\lhat$) for the $n=4$ structure as opposed to polarization conversion for the $n=1$ structure. Again, these findings do not depend on the number of helix turns and loss model. These results are consistent with the result in Eq. (\ref{eq:back}): Helicity cannot be preserved in backward scattering when $n\ge 3$. In backward scattering, helicity change translates into preservation of the real space circular polarization vectors (see Table \ref{tab:helreal}).

One of the properties of the symmetry arguments that I have used in the derivations is that they are independent of factors like the wavelength of the illumination, the number of turns of the helices, the loss model, or whether the system is an array, a square arrangement of four helices or a single helix. 

The restrictions on the forward and backward helicity scattering coefficients for $n\ge 3$ agree with the results in \cite{Hu1987}. In that work, the authors study the consequences that different geometrical symmetries of the scatterer, including discrete rotational symmetries, have on the forward and backward scattering Mueller and Jones matrices. 

Before involving the duality symmetry in the discussion, it is worth considering the results in \cite{Hopkins2013}. The authors show that the extinction, absorption and scattering cross sections of nanoparticle clusters with $n\ge 3$ discrete rotational symmetry are independent of the linear polarization angle of the input plane wave. Using the formalism of this paper, the polarization independence of the extinction cross section can be recovered for systems with $n\ge 3$ featuring a mirror plane of symmetry ($M$) containing the symmetry axis. For a given input polarization, the optical theorem \cite[Chap. 10.11]{Jackson1998} states that the total extinction cross section is proportional to the imaginary part of the co-polarization forward scattering coefficient. Let me show that the mirror symmetry forces the forward scattering coefficient to be identical for both helicities:

\begin{equation}
\label{eq:ref}
\tau_f^{\lambda\lambda}=\langle \lambda \ p\zhat|S|p\zhat\ \lambda\rangle=\langle  \lambda \ p\zhat|M^{\dagger}SM|p\zhat \ \lambda\rangle=\langle -\lambda \ p\zhat|S|p\zhat \ -\lambda\rangle=\tau_f^{-\lambda-\lambda},
\end{equation}
where the second equality follows from the invariance of the scatterer under the mirror reflection, and the third one from the fact that the momentum $p\zhat$ is left unchanged since the mirror plane contains the $z$ axis, but helicity flips sign under any spatial inversion.

Together with the inherent helicity preservation of the $n\ge3$ system, which means that $\tau_f^{\lambda,-\lambda}=0$, Eq. (\ref{eq:ref}) implies that, in the helicity basis, the 2x2 Jones matrix is proportional to the identity. It is therefore also proportional to the identity in all polarization basis obtained from the helicity basis by a unitary transformation, in particular, in the linear polarization basis. The extinction cross section will hence be independent of the polarization. It is interesting to note that, while the structures considered in \cite{Hopkins2013} are mirror symmetric, the ones in \cite{Decker2010} and \cite{Kaschke2012} are not, so the above conclusion does not apply to them.

Consider now the following question: What happens if, on top of a discrete rotational symmetry with $n\ge3$, the scatterer has duality symmetry? In this case, there will not be any scattering in the backwards direction at all:  Since $n\ge3$, only the changed helicity component is possible, but, because of duality symmetry (helicity preservation) helicity cannot change upon scattering. The solution is that $\tau_b^{\lambda\bar{\lambda}}=0$ for all $(\lambda,\bar{\lambda})$. The system exhibits zero backscattering. 

Zero backscattering from dual systems with $R_z(2\pi/4)$ rotation has been studied in detail \cite{Lindell2009,Karilainen2012}. In this article, the consideration of the connection between helicity and duality, and the relationship between helicity and angular momentum in this restricted case, allows to conclude that the same zero backscattering effect exists for $R_z(2\pi/3),R_z(2\pi/5),R_z(2\pi/6),...$ symmetric scatterers. When $n\rightarrow \infty$, the system reaches the symmetry of a cylinder. A particular case of a cylindrically symmetric dual object exhibiting zero backscattering was studied by Kerker \cite{Kerker1983} in his work about magnetic spheres with $\epsilon=\mu$. These spheres have duality symmetry when immersed in vacuum \cite{FerCor2012p}. Zero backscattering can be generalized beyond spheres to any dual and cylindrically symmetric object \cite{Zambrana2013}, and, as shown in this paper, to systems with discrete rotational symmetries.

To finalize, let me assume that a system has the property of always flipping the helicity of a general field interacting with it, that is, it has an anti-dual behavior \cite{Zambrana2013b}. It can be proved that an anti-dual system with a discrete rotational symmetry of $n\ge3$ would exhibit zero forward scattering. This can be seen graphically in Fig. \ref{fig:nyaca} by imagining the removal of all the scattered plane waves with helicity equal to the input one. Zero forward scattering has drastic implications regarding the optical theorem, which relates the total scattered and absorbed power to the forward scattering coefficients. This conflict between zero forward scattering and the optical theorem has been recognized and studied in detail for spherical dielectric scatterers \cite{Alu2010}. For a passive anti-dual object with $n\ge 3$, the optical theorem implies that such object would be totally transparent to electromagnetic radiation (null scattering operator $S=0$). Anti-dual objects, if they exist, must then be made of active materials \cite{Zambrana2013b}. Nevertheless, mode dependent approximate anti-dual behavior can be observed in spherical dielectric spheres \cite{Zambrana2013b}.

A direct application of the results contained in this paper is the design of a planar array of dual symmetric inclusions ordered so that the system has a discrete rotational symmetry of degree $n\ge 3$. Such structure should exhibit zero backscattering. One possible kind of dual symmetric inclusions are dielectric spheres \cite{Zambrana2013b}: When they have the appropriate relationship between size and dielectric constant, they become dual symmetric to a very good approximation. Recently, solar cells of semiconductor nanowires arranged in a square lattice have been shown to achieve significant efficiencies \cite{Wallentin2013}. According to the results of this paper, investigating the duality properties of the nanowires could lead to insights for reducing their reflection of normally incident light.

In conclusion, I have used the framework of symmetries and conservation laws to analyze the forward and backward scattering properties of systems with discrete rotational symmetries $R_z(2\pi/n)$. The analysis shows that the helicity scattering coefficients are restricted for systems with symmetries of degree $n\ge3$: Forward scattering can only be helicity preserving while backward scattering can only be helicity flipping. No such restrictions exists for systems of degree $n=2$, or the trivial $n=1$. If, in addition to the discrete rotational symmetry of degree $n\ge3$, the system has electromagnetic duality symmetry, it will exhibit zero backscattering. The results pinpoint the underlying symmetry reasons for some notable scattering properties of $R_z(2\pi/4)$ symmetric systems that have been reported in the metamaterials \cite{Decker2010,Kaschke2012} and radar literature \cite{Lindell2009,Karilainen2012}. This paper shows that systems with an $R_z(2\pi/n)$ symmetry of degree $n\ge 3$ exhibit the same effects. For example, a planar array of dual symmetric inclusions designed with an $n\ge3$ symmetry will exhibit zero backscattering. The results may be relevant for reducing the reflectivity of some kinds of solar cells.

\noindent {\bf Acknowledgments}

\noindent I wish to acknowledge useful conversations with Dr. Mathieu Juan and A. Prof. Gabriel Molina-Terriza. This work was funded by an iMQRES scholarship of Macquarie University, and the Centre of Excellence for Engineered Quantum Systems (EQuS).
%\bibliography{ivanfcrefs}%RT
\end{document}